\documentclass[a4paper, 12pt]{article} 
\usepackage[right=1.1in,left=1.1in,top=1.1in,bottom=1.1in]{geometry}
\usepackage{graphicx} 
\usepackage{braket}
\usepackage{mathrsfs}
\usepackage[T1]{fontenc} 
\usepackage{amsmath}
\usepackage{pxfonts}
\usepackage[T1]{fontenc}
\usepackage{amssymb}
\usepackage{epstopdf}
\usepackage{setspace}
\usepackage{enumitem}
\usepackage[ comma, authoryear]{natbib}
\usepackage{sectsty}
\usepackage{wrapfig} 
\sectionfont{\large}
\bibpunct{(}{)}{,}{a}{}{,}
\usepackage{tikz}   
\usepackage{tikz-3dplot} 

\newtheorem{theorem}{Theorem}[section]

\newcommand{\qed}{\nobreak \ifvmode \relax \else
      \ifdim\lastskip<1.5em \hskip-\lastskip
      \hskip1.5em plus0em minus0.5em \fi \nobreak
      \vrule height0.75em width0.5em depth0.25em\fi}

\usepackage{bbm}
\usepackage{MnSymbol}
\usepackage[all]{xy}

\usepackage{xcolor,colortbl,array,amssymb}

\makeatletter

\title{The Intrinsic Structure of Quantum Mechanics} 

\author{Eddy Keming Chen\thanks{Department of Philosophy, 106 Somerset Street, Rutgers University, New Brunswick, NJ 08901, USA. Website: www.eddykemingchen.net. Email: eddy.chen@rutgers.edu  }}  

\date{\today} 

\begin{document}
\bibliographystyle{plain}

\maketitle 



\begin{abstract}

The wave function in quantum mechanics presents an interesting challenge to our understanding of the physical world.
In this paper, I show that the wave function can be understood as four intrinsic relations on physical space. My account has three desirable features that  the standard account lacks: (1) it does not refer to any abstract mathematical objects, (2) it is free from the usual arbitrary conventions, and (3) it explains why the wave function has its gauge degrees of freedom, something that are usually put into the theory by hand. 

Hence, this account has  implications for debates in philosophy of mathematics and philosophy of science. First, by removing references to mathematical objects, it provides a framework for nominalizing quantum mechanics. Second, by excising superfluous structure such as overall phase, it reveals the intrinsic structure postulated by quantum mechanics. Moreover, it also removes a major obstacle to ``wave function realism.'' 


\end{abstract}

\hspace*{3,6mm}\textit{Keywords: quantum mechanics, wave function, structural realism, phase structure, mathematical nominalism vs. platonism, foundations of measurement, intrinsic physical theory, Quine-Putnam indispensability argument, metaphysics of science. }   

\newpage

\begingroup
\singlespacing
\tableofcontents
\endgroup

\vspace{20pt} 







\nocite{FieldSWN, durr2012quantum, bell2004speakable, forrest1988quantum, belot2012quantum, ney2013wave, ArntzeniusDorrCG, ChenOurFund, field2016science, AlbertEQM, allori2013primitive, north2009structure, dasgupta2013absolutism, martens2016against, baker2014some, sep-nominalism-mathematics, bueno2003possible}

\section{Introduction}

Quantum mechanics is empirically successful (at least in the non-relativistic domain). But what it means remains highly controversial. Since its initial formulation, there have been many debates (in physics and in philosophy) about the ontology of a quantum-mechanical world. Chief among them is a serious foundational question about how  to understand the quantum-mechanical laws and the origin of quantum randomness.  That is the topic of the quantum measurement problem.
At the time of writing this paper, the following are  serious contenders for being the best solution: Bohmian mechanics (BM), spontaneous localization theories (GRW0, GRWf, GRWm, CSL), and Everettian quantum mechanics (EQM and Many-Worlds Interpretation (MWI)).\footnote{See \cite{norsen2017} for an updated introduction to the measurement problem and the main solutions.} 

There are other deep questions about quantum mechanics that have a philosophical and metaphysical flavor. Opening a standard textbook on quantum mechanics, we find an abundance of mathematical objects: Hilbert spaces, operators, matrices, wave functions, and etc. But what do they represent in the physical world? Are they ontologically serious to the same degree or are some merely dispensable instruments that facilitate calculations? In recent debates in metaphysics of quantum mechanics, there is considerable agreement that the universal wave function, modulo some mathematical degrees of freedom, represents something  objective --- the quantum state of the universe.\footnote{The universal quantum state, represented by a universal wave function, can give rise to wave functions of the subsystems. The clearest examples are the conditional wave functions in Bohmian mechanics. However, our primary focus here will be on the wave function of the universe.}  In contrast, matrices and operators are merely convenient summaries that do not play the same fundamental role as the wave function. 

However, the meaning of the universal quantum state is far from clear. We know its mathematical representation very well: the universal wave function, which is crucially involved in the dynamics of BM, GRW, and EQM. In the position representation, a scalar-valued wave function is a square-integrable function from the configuration space $\mathbb{R}^{3N}$ to the complex plane $\mathbb{C}$.  But what does the wave function really mean? There are two ways of pursuing this question:

\begin{enumerate}
\item What kind of ``thing'' does the wave function represent? Does it represent a physical field on the configuration space, something nomological, or a \emph{sui generis} entity in its own ontological category? 
\item What is the physical basis for the mathematics used for the wave function? Which mathematical degrees of freedom of the wave function are physically genuine? What is the metaphysical explanation for the merely mathematical or gauge degrees of freedom?  
\end{enumerate}

Much of the philosophical literature on the metaphysics of the wave function has pursued the first line of questions.\footnote{See, for example, \cite{AlbertEQM, LoewerHS, wallace2010quantum, NorthSQW, NeySOTDQU, maudlin2013nature, goldstein2013reality, miller2013quantum, bhogal2015humean}.} In this paper, I will primarily pursue the second one, but I will also show that these two are intimately related.

In particular, I will introduce an intrinsic account of the quantum state.  It answers the second line of questions by picking out four concrete relations on physical space-time. Thus, it makes explicit the physical basis for the usefulness of the mathematics of the wave function,  and it provides a metaphysical explanation for why certain degrees of freedom in the wave function (the scale of the amplitude and the overall phase) are merely gauge. The intrinsic account  also has the feature that the fundamental ontology does not include abstract mathematical objects such as complex numbers, functions, vectors, or sets. 

The intrinsic account is therefore nominalistic in the sense of Hartry Field (1980). In his influential monograph \emph{Science Without Numbers: A Defense of Nominalism}, Field advances a new approach to philosophy of mathematics by explicitly constructing nominalistic counterparts of the platonistic physical theories. In particular, he nominalizes Newtonian gravitation theory.\footnote{It is not quite complete as it leaves out integration. } In the same spirit, Frank Arntzenius and Cian Dorr (2011) develop a nominalization of differential manifolds, laying down the foundation of a nominalistic theory of classical field theories and general relativity. Up until now, however, there has been no successful nominalization of quantum theory. In fact, it has been an open problem--both conceptually and mathematically--how it is to be done. The non-existence of a nominalistic quantum mechanics has encouraged much skepticism about Field's program of nominalizing fundamental physics and much optimism about the Quine-Putnam Indispensability Argument for Mathematical Objects. Indeed, there is a long-standing conjecture, due to David Malament (1982), that Field's nominalism would not succeed in quantum mechanics.  Therefore, being nominalistic, my intrinsic theory of the quantum state would advance Field's nominalistic project and provide (the first step of) an answer to Malament's skepticism. 

Another interesting consequence of the account is that  it will make progress on the first line of questions about the ontology of quantum mechanics. On an increasingly influential interpretation of the wave function,  it represents something physically significant. One version of this view is the so-called ``wave function realism,'' the view that the universal wave function represents a physical field on a high-dimensional (fundamental) space. That is the position developed and defended in Albert (1996),  Loewer (1996),  Ney (2012), and  North (2013). However, Tim Maudlin (2013) has argued that  this view leads to an unpleasant proliferation of possibilities: if the wave function represents a physical field (like the classical electromagnetic field), then a change of the wave function by an overall phase transformation will produce a distinct physical possibility. But the two wave functions will be empirically equivalent---no experiments can distinguish them, which is the reason why the overall phase differences are usually regarded as merely \emph{gauge}. Since the intrinsic account of the wave function I offer here is gauge-free insofar as overall phase is concerned, it  removes a major obstacle to wave function realism (vis-\`a-vis Maudlin's objection). 

In this paper, I will first explain (in \S 2) the two visions for a fundamental physical theory of the world: the intrinsicalist vision and the nominalistic vision. I will then discuss why quantum theory may seem to resist the intrinsic and nominalistic reformulation. Next (in \S 3), I will write down an intrinsic and nominalistic theory of the quantum state. 
Finally (in \S 4), I will discuss how this account bears on the nature of phase and the debate about \emph{wave function realism}. 


Along the way, I axiomatize the quantum phase structure as what I shall call a \emph{periodic difference structure}  and prove a representation theorem and a uniqueness theorem. These formal results could prove fruitful for further investigation into the metaphysics of quantum mechanics and theoretical structure in physical theories.

\section{The Two Visions and the Quantum Obstacle}

There are, broadly speaking, two grand visions for what a fundamental physical theory of the world should look like. (To be sure, there are many other visions and aspirations.) The first is what I shall call the intrinsicalist vision, the requirement that the fundamental theory be written in a form without any reference to arbitrary conventions such as coordinate systems and units of scale. The second is the nominalistic vision, the requirement that the fundamental theory be written without any reference to mathematical objects. The first one is familiar to mathematical physicists from the development of synthetic geometry and differential geometry. The second one is familiar to philosophers of mathematics and philosophers of physics working on the ontological commitment of physical theories. First, I will describe the two visions, explain their motivations, and provide some examples. Next, I will  explain why quantum mechanics seems to be an obstacle for both programs. 

\subsection{The Intrinsicalist Vision}

\begin{figure}
\center
\includegraphics[scale=0.5]{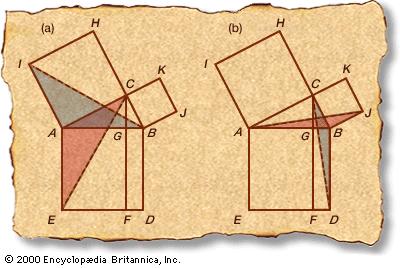}
\caption{ Euclid's Windmill proof of the Pythagorean Theorem. No coordinate systems or real numbers were used. Cartesian coordinates were invented much later to facilitate derivations.}
\end{figure}

The intrinsicalist vision is best illustrated with some history of Euclidean geometry. Euclid showed that complex geometrical facts can be demonstrated using rigorous proof on the basis of simple axioms. However, Euclid's axioms do not mention real numbers or coordinate systems, for they were not yet discovered. They are stated with only qualitative predicates such as the equality of line segments and the congruence of angles.  With these concepts, Euclid was able to derive a large body of geometrical propositions.  

Real numbers and coordinate systems were introduced to facilitate the derivations. With the full power of real analysis, the metric function defined on pairs of tuples of coordinate numbers can greatly speed up the calculations, which usually take up many steps of logical derivation on Euclid's approach.   But what are the significance of the real numbers and coordinate systems? When representing a 3-dimensional Euclidean space, a typical choice is to use $\mathbb{R}^3$. It is clear that such a representation has much surplus (or excess) structure: the origin of the coordinate system, the orientation of the axis, and the scale are all arbitrarily chosen (sometimes conveniently chosen for ease of calculation). There is ``more information'' or ``more structure'' in $\mathbb{R}^3$ than in the 3-dimensional Euclidean space. In other words, the $\mathbb{R}^3$ representation has gauge degrees of freedom. 

The real, intrinsic structure in the 3-dimensional Euclidean space--the structure that is represented by $\mathbb{R}^3$ up to the Euclidean transformations--can be understood as an axiomatic structure of congruence and betweenness. In fact, Hilbert 1899 and Tarski 1959 give us ways to make this statement more precise.  After offering a rigorous axiomatization of Euclidean geometry, they prove a representation theorem: any structure instantiates the betweenness and congruence axioms of 3-dimensional Euclidean geometry if and only if there is a 1-1 embedding function from the structure onto $\mathbb{R}^3$ such that if we define a metric function in the usual Pythagorean way then the metric function is homomorphic: it preserves the exact structure of betweenness and congruence. Moreover, they prove a uniqueness theorem: any other embedding function defined on the same domain satisfies the same conditions of homomorphism if and only if it is a Euclidean transformation of the original embedding function: a transformation on $\mathbb{R}^3$ that can be obtained by some combination of shift of origin, reflection, rotation, and positive scaling. 

The formal results support the idea that we can think of the genuine, intrinsic features of 3-dimensional Euclidean space as consisting directly of betweenness and congruence relations on spatial points, and we can regard the coordinate system ($\mathbb{R}^3$) and the metric function as extrinsic representational features we bring to facilitate calculations. (Example: Figure 1. Exercise: prove the Pythagorean Theorem with and without real-numbered coordinate systems.) The merely representational artifacts are highly useful but still dispensable.

There are several advantages of having an intrinsic formulation of geometry. First, it eliminates the need for many arbitrary conventions: where to place the origin, how to orient the axis, and what scale to use. Second, in the absence of these arbitrary conventions, we have a theory whose elements \emph{could} stand in one-to-one correspondence with elements of reality. In that case, we can look directly into the real structure of the geometrical objects without worrying that we are looking at some merely representational artifact (or gauge degrees of freedom). By eliminating redundant structure in a theory, an intrinsic formulation gives us a more perspicuous picture of the geometrical reality. 

The lessons we learn from the history of Euclidean geometry can be extended to other parts of physics. For example, people have long noticed that there are many gauge degrees of freedom in the representation of both scalar and vector valued physical quantities: temperature, mass, potential, and field values. There has been much debate in philosophy of physics about what structure is physically genuine and and what is merely gauge. It would therefore be helpful to go beyond the scope of physical geometry and extend the intrinsic approach to physical theories in general. 

Hartry Field (1980), building on previous work by \cite{KrantzLuceSuppesTversky}, ingeniously extends the intrinsic approach to Newtonian gravitation theory. The result is an elimination of arbitrary choices of zero  field value and units of mass. His conjecture is that all physical theories can be ``intrinsicalized'' in one way or another.

\subsection{The Nominalist Vision}

As mentioned earlier, Field (1980) provides an intrinsic version of Newtonian gravitation theory. But the main motivation and the major achievement of his project is a defense of nominalism, the thesis that there are no abstract entities, and, in particular, no abstract mathematical entities such as numbers, functions, and sets. 

The background for Field's nominalistic project is the classic debate between the mathematical nominalist and the mathematical platonist, the latter of whom is ontologically committed to the existence of abstract mathematical objects. Field identifies a main problem of maintaining nominalism is the apparent indispensability of mathematical objects in formulating our best physical theories:

\begin{quote}
\text{          } Since I deny that numbers, functions, sets, etc. exist, I deny that it is legitimate to use terms that purport to refer to such entities, or variables that purport to range over such entities, in our ultimate account of what the world is really like.

\text{          } This appears to raise a problem: for our ultimate account of what the world is really like must surely include a physical theory; and in developing physical theories one needs to use mathematics; and mathematics is full of such references to and quantifications over numbers, functions, sets, and the like. It would appear then that nominalism is not a position that can reasonably be maintained.\footnote{Field (2016), Preliminary Remarks, p.1.}
\end{quote}
In other words, the main task of defending nominalism would be to respond to the Quine-Putnam Indispensability Argument:\footnote{The argument was originally proposed by W. V. Quine and later developed by \cite{PutnamPL}. This version is from \cite{sep-mathphil-indis}. }

\begin{enumerate}
\item[P1] We ought to be ontologically committed to all (and only) those entities that are indispensable to our best theories of the world. [Quine's Criterion of Ontological Commitment]
\item[P2] Mathematical entities are indispensable to our best theories of the world. [The Indispensability Thesis]
\item[C] Therefore, we ought to be ontologically committed to mathematical entities. 
\end{enumerate}

In particular, Field's task is to refute the second premise--the Indispensability Thesis. Field proposes to replace all platonistic physical theories with attractive nominalistic versions that do not quantify over mathematical objects

Field's nominalistic versions of physical theories would have significant advantages over their platonistic counterparts. First, the nominalistic versions illuminate what exactly in the physical world provide the explanations for the usefulness of any particular mathematical representation. After all, even  a platonist might accept that numbers and coordinate systems do not really exist in the physical world but merely represent some concrete physical reality. Such an attitude is consistent with the platonist's endorsement of the Indispensability Thesis. Second, as Field has argued, the nominalistic physics seems to provide better explanations than the platonistic counterparts, for the latter would involve explanation of physical phenomena by things (such as numbers) external to the physical processes themselves. 

Field has partially succeeded by writing down an intrinsic theory of physical geometry and Newtonian gravitation, as it contains no explicit first-order quantification over mathematical objects, thus qualifying his theory as nominalistic. But what about other theories? Despite the initial success of his project, there has been significant skepticism about whether his project can extend beyond Newtonian gravitation theory to more advanced theories such as quantum mechanics.

\subsection{Obstacles From Quantum Theory}

We have looked at the motivations for the two visions for what the fundamental theory of the world should look like: the intrinsicalist vision and the nominalistic vision. They should not be thought of as competing against each other. They often converge on a common project. Indeed, Field's reformulation of Newtonian Gravitation Theory is both intrinsic and nominalistic.\footnote{However, the intrinsicalist and nominalistic visions can also come apart. For example, we can, in the case of mass, adopt an intrinsic yet platonistic theory of mass ratios. We can also adopt an extrinsic yet nominalistic theory of mass relations by using some arbitrary object (say, my water bottle) as standing for unit mass and assigning comparative relations between that arbitrary object and every other object.} 

Both have had considerable success in certain segments of classical theories. But with the rich mathematical structures and abstract formalisms in quantum mechanics, both seem to run into obstacles. David Malament  was one of the earliest critics of the nominalistic vision.  He voiced his skepticism in his influential review of Field's book. Malament states his general worry as follows:

\begin{quote}
\text{  } Suppose Field wants to give some physical theory a nominalistic reformulation. Further suppose the theory determines a class of mathematical models, each of which consists of a set of ``points'' together with certain mathematical structures defined on them. Field's nominalization strategy cannot be successful unless the objects represented by the points are appropriately physical (or non-abstract)...But in lots of cases the represented objects \emph{are} abstract. (\cite{MalamentFR}, pp. 533, emphasis original.)\footnote{Malament also gives the example of classical Hamiltonian mechanics as another specific instance of the general worry. But this is not the place to get into classical mechanics. Suffice to say that there are several ways to nominalize classical mechanics. Field's nominalistic Newtonian Gravitation Theory is one way. \cite{ArntzeniusDorrCG} provides another way.}  
\end{quote}
Given his \emph{general} worry that, often in physical theories, it is abstracta that are represented in the state spaces, Malament conjectures that, in the specific case of quantum mechanics, Field's strategy of nominalization would not ``have a chance'': 

\begin{quote}
\text{  } Here [in the context of quantum mechanics] I do not really see how Field can get started at all. I suppose one can think of the theory as determining a set of models---each a Hilbert space. But what form would the recovery (i.e., representation) theorem take? The only possibility that comes to mind is a theorem of the sort sought by Jauch, Piron, \emph{et al}. They start with ``propositions'' (or ``eventualities'') and lattice-theoretic relations as primitive, and then seek to prove that the lattice of propositions is necessarily isomorphic to the lattice of subspaces of some Hilbert space. But of course no theorem of this sort would be of any use to Field. What could be worse than \emph{propositions} (or \emph{eventualities})? (\cite{MalamentFR}, pp. 533-34.)
\end{quote}
As I understand it, Malament suggests that there are no good places to start nominalizing non-relativistic quantum mechanics. This is because the obvious starting point, according to Malament and other commentators, is the abstract Hilbert space, $\mathscr{H}$, as it is a state space of the quantum state. 

However, there may be other starting points to nominalize quantum mechanics. For example, the configuration space, $\mathbb{R}^{3N}$, is a good candidate. In realist quantum theories such as Bohmian mechanics, Everettian quantum mechanics, and spontaneous localization theories, it is standard to postulate a (normalized) universal wave function $\Psi(\textbf{x},t)$ defined on the configuration space(-time) and a dynamical equation governing its temporal evolution.\footnote{Bohmian mechanics postulates additional ontologies---particles with precise locations in physical space---and an extra law of motion---the guidance equation. GRW theories postulate an additional stochastic modification of the Schr\"odinger equation and, for some versions, additional ontologies such as flashes and mass densities in physical space.} In the deterministic case, the wave function evolves according to the Schr\"odinger equation, 
$$i\hbar \frac{\partial}{\partial t} \Psi(\textbf{x}, t) = [- \sum^{N}_{i=1} \frac{\hbar^2}{2m_i} \Delta_i + V(x)] \Psi(\textbf{x},t) := H \Psi(\textbf{x},t),$$
which relates the temporal derivatives of the wave function to its spatial derivatives. Now, the configuration-space viewpoint can be translated into the Hilbert space formalism. If we regard the wave function (a square-integrable function from the configuration space to complex numbers) as a unit vector $\ket{\Psi(t)}$, then we can form another space---the  Hilbert space of the system.\footnote{This is the Hilbert space $L^2 (\mathbb{R}^{3N}, \mathbb{C})$, equipped with the inner product $<\psi, \phi>$ of taking the Lebesgue integral of $\psi^{*} \phi$ over the configuration space, which guarantees Cauchy Completeness. } Thus, the wave function can be mapped to a state vector, and vice versa. The state vector then rotates (on the unit sphere in the Hilbert space) according to a unitary (Hamiltonian) operator, 
$$i\hbar  \frac{\partial}{\partial t} \ket{\Psi(t)} = \hat{H} \ket{\Psi(t)} ,$$
which is another way to express the Schr\"odinger evolution of the wave function.

Hence, there is the possibility of carrying out the nominalization project with the configuration space. In some respects, the configuration-space viewpoint is  more friendly to  nominalism, as the configuration space is much closely related to  physical space than the abstract Hilbert space is.\footnote{I should emphasize that, because of its central role in functional analysis, Hilbert space is highly important for fascilitating calculations and proving theorems about quantum mechanics. Nevertheless, we should not regard it as conclusive evidence for ontological priority. Indeed, as we shall see in \S 3, the configuration-space viewpoint provides a natural platform for the nominalization of the universal wave function. We should also keep in mind that, at the end of the day, it suffices to show that quantum mechanics can be successfully nominalized from \emph{some} viewpoint.} Nevertheless, Malament's worries still remain, because  (\emph{prima facie}) the configuration space is also quite abstract, and it is unclear how to fit it into the nominalistic framework. Therefore, at least \emph{prima facie}, quantum mechanics seems to frustrate the nominalistic vision. 


Moreover, the mathematics of quantum mechanics comes with much conventional structure that is hard to get rid of. For example, we know that the exact value of the amplitude of the wave function is not important. For that matter, we can scale it with any arbitrary positive constant. It is true that we usually choose the scale such that we get unity when integrating the amplitude over the entire configuration space. But that is merely conventional. We can, for example, write down the Born rule with a proportionality constant to get unity in the probability function:

$$P(x\in X) = Z \int_X |\Psi(x)|^2 dx,$$
where $Z$ is a normalization constant. 

Another example is the overall phase of the wave function. As we learn from modular arithmetic, the exact value of the phase of the wave function is not physically significant, as we can add a constant phase factor to every point in configuration space and the wave function will remain physically the same: producing exactly the same predictions in terms of probabilities. 

All these gauge degrees of freedom are frustrating from the point of view of the intrinsicalist vision. They are the manifestation of excess structures in the quantum theory. What exactly is going on in the real world that allows for these gauge degrees of freedom but not others? What is the most metaphysically perspicuous picture of the quantum state, represented by the wave function? Many people would respond that the quantum state is projective, meaning that the state space for the quantum state is not the Hilbert space, but its quotient space: the projective Hilbert space. It can be obtained by quotienting the usual Hilbert space with the equivalence relation $ \psi \sim Re^{i\theta}\psi $. But this is not satisfying; the ``quotienting'' strategy raises a similar question: what exactly is going on in the real world that allows for quotienting with this equivalence relation but not others?\footnote{These questions, I believe, are in the same spirit as Ted Sider's 2016 Locke Lecture (ms.), and especially his final lecture on theoretical equivalence and against what he calls``quotienting by hand.'' I should mention that both Sider and I are really after gauge-free formulations of physical and metaphysical theories, which are  more stringent than merely gauge-independent formulations. For example, modern differential geometry is gauge-independent (coordinate-independent) but not gauge-free (coordinate-free): although manifolds can be defined without privileging any particular coordinate system, their definition still uses coordinate systems (maps and atlas).} No one, as far as I know, has offered an intrinsic picture of the quantum state, even in the non-relativistic domain. 

In short, at least \emph{prima facie}, both the intrinsicalist vision and the nominalist vision are challenged by quantum mechanics. 




\section{An Intrinsic and Nominalistic Account of the Quantum State}

In this section, I  propose a new account of the quantum state based on some lessons we learned from the debates about wave function realism.\footnote{Here I'm taking the ``Hard Road'' to nominalism. As such, my goal is to (1) reformulate quantum mechanics (QM) such that \emph{within the theory} it no longer refers (under first-order quantifiers) to mathematical objects such as numbers, functions, or sets and (2) demonstrate that the platonistic version of QM is conservative over the nominalistic reformulation. To arrive at my theory, and to state and prove the representation theorems, I refer to some mathematical objects. But these are parts of the meta-theory to explicate the relation between my account and the platonistic counterpart and to argue (by \emph{reductio}) against the indispensability thesis. See Field (2016), Preliminary Remarks and Ch. 1 for a clear discussion, and \cite{colyvan2010there} for an up-to-date assessment of the ``Easy Road'' option. Thanks to Andrea Oldofredi and Ted Sider for suggesting that I make this clear.} As we shall see, it does not take much to overcome the ``quantum obstacle.''  For simplicity, I will focus on the case of a quantum state for a constant number of identical particles without spin.

\subsection{The Mathematics of the Quantum State}

First, let me explain my strategy for nominalizing non-relativistic quantum mechanics. 
\begin{enumerate}
\item I will start with a Newtonian space-time, whose nominalization is readily available.\footnote{It is an interesting question what role Galilean relativity plays in non-relativistic quantum mechanics. I will explore this issue in future work.} 
\item I will use symmetries as a guide to fundamentality and identify the intrinsic structure of the universal quantum state  on the Newtonian space-time. This will be the goal for the remaining part of the paper. (Here we focus only on the quantum state, because it is novel and it seems to resist nominalization. But the theory leaves room for additional ontologies of particles, fields, mass densities supplied by specific interpretations of QM; these additional ontologies are readily nominalizable.)
\item In future work, I will develop nominalistic translations of the dynamical equations and generalize this account to accommodate more complicated quantum theories. 
\end{enumerate}

Before we get into the intrinsic structure of the universal quantum state, we need to say a bit more about its mathematical structure. For the quantum state of a spinless system at a time $t$ (see Figure 2), we can represent it with a scalar-valued wave function: 

$$\Psi_t : \mathbb{R}^{3N} \rightarrow \mathbb{C},$$
where $N$ is the number of particles in the system, $\mathbb{R}^{3N}$ is the configuration space of $N$ particles, and $ \mathbb{C}$  is the complex plane. (For the quantum state of a system with spin, we can use a vector-valued wave function whose range is the spinor space---$\mathbb{C}^{2^N}$.) 

My strategy is to start with a Newtonian space-time (which is usually represented by a Cartesian product of a 3-dimensional Euclidean space and a 1-dimensional time). If we want to nominalize the quantum state, what should we do with the configuration space $ \mathbb{R}^{3N} $? As is now familiar from the debate about wave function realism, there are two ways of interpreting the fundamental physical space for a quantum world:

\begin{figure}
\center
\includegraphics[scale=0.45]{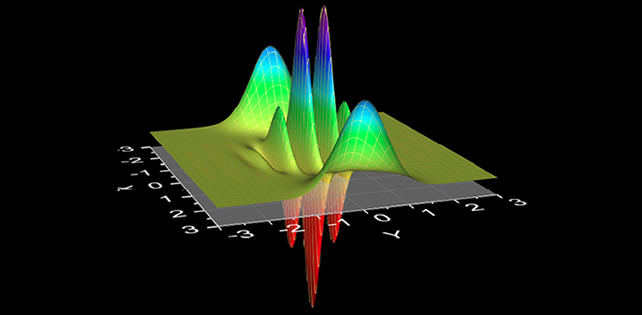}
\caption{A wave function for a two-particle system in 1-dim space. Source: www.physics.auckland.ac.nz}
\end{figure}

\begin{enumerate}
\item $\mathbb{R}^{3N}$ represents the fundamental physical space; the space represented by $\mathbb{R}^{3}$ only \emph{appears} to be real; the quantum state assigns a complex number to each point in $\mathbb{R}^{3N}$. (See Figure 2. Analogy: classical field.)
\item $\mathbb{R}^{3}$ represents the fundamental physical space; the space represented by $\mathbb{R}^{3N}$ is a mathematical construction---the configuration space; the quantum state assigns a complex number to each region in $\mathbb{R}^{3}$ that contains $N$ points (i.e. the regions will be irregular and disconnected). (Analogy: multi-field)
\end{enumerate}

Some authors in the debate about wave function realism have argued that given our current total evidence, option (2) is a  better interpretation of non-relativistic quantum mechanics.\footnote{See, for example, \cite{ChenOurFund} and \cite{Hubert2018}.} I will not rehearse their arguments here. But one of the key ideas that will help us here is that we can think of the complex-valued function as really ``living on'' the 3-dimensional physical space, in the sense that it assigns a complex number not to each \emph{point} but each $N$-element \emph{region} in physical space. We call that a ``multi-field.''\footnote{This name can be a little confusing. Wave-function ``multi-field'' was first used in \cite{belot2012quantum}, which was an adaptation of the name ``polyfield'' introduced by Forrest (1988). See  \cite{ArntzeniusDorrCG} for a completely different object called the ``multi-field.'' } 

Taking the wave function into a framework friendly for further nominalization, we can perform the familiar technique of decomposing the complex number $Re^{i\theta}$ into two real numbers: the amplitude $R$ and the phase $\theta$. That is, we can think of the compex-valued  multi-field in the physical space as two real-valued multi-fields:

$$R(x_{1}, x_{2}, x_{3}, ... , x_{N}), \theta(x_{1}, x_{2}, x_{3}, ... , x_{N}).$$ 
Here, since we are discussing Newtonian space-time, the $x_1..... x_N$ are simultaneous space-time points. We can think of them as:   $(x_{\alpha_1}, x_{\beta_1}, x_{\gamma_1}, x_{t}),$ $ (x_{\alpha_2}, x_{\beta_2}, x_{\gamma_2}, x_{t}),......,$ $(x_{\alpha_N}, x_{\beta_N}, x_{\gamma_N}, x_{t})$. 

Now the task before us is just to come up with a nominalistic and intrinsic description of the two multi-fields. In \S 3.2 and \S 3.3, we will find two physical structures (Quantum State Amplitude and Quantum State Phase), which, via the appropriate representation theorems and uniqueness theorems, justify the use of complex numbers and explain the gauge degrees of freedom in the quantum wave function.\footnote{In the case of a vector-valued wave function, since the wave function value consists in $2^N$ complex numbers, where $N$ is the number of particles, we would need to nominalize $2^{N+1}$ real-valued functions: $$R_1 (x_{1}, x_{2}, x_{3}, ... , x_{N}), \theta_1 (x_{1}, x_{2}, x_{3}, ... , x_{N}), R_2 (x_{1}, x_{2}, x_{3}, ... , x_{N}), \theta_2 (x_{1}, x_{2}, x_{3}, ... , x_{N}),......$$ 
}

\subsection{Quantum State Amplitude}

The amplitude part of the quantum state is (like mass density) on the ratio scale, i.e. the physical structure should be invariant under ratio transformations $$R \rightarrow \alpha R.$$

We will start with the Newtonian space-time and help ourselves to the structure of \textbf{N-Regions}: collection of all  regions that contain exactly $N$ simultaneous space-time points  (which are irregular and disconnected regions). We start here because we would like to have a physical realization of the platonistic configuration space. The solution is to identify configuration points with certain special regions of the physical space-time.\footnote{Notes on mereology: As I am taking for granted that quantum mechanics for indistinguishable particles (sometimes called identical particles) works just as well as quantum mechanics for distinguishable particles, I do not require anything more than Atomistic General Extensional Mereology (AGEM). That is, the mereological system that validate the following principles: Partial Ordering of Parthood, Strong Supplementation, Unrestricted Fusion, and Atomicity.  See \cite{sep-mereology} for a detailed discussion. 

However, I leave open the possibility for adding structures in \textbf{N-Regions} to distinguish among different ways of forming regions from the same collection of points, corresponding to permuted configurations of distinguishable particles. We might need to introduce additional structure for mereological composition to distinguish between mereological sums formed from the same atoms but in different orders. This might also be required when we have entangled quantum states of different species of particles. To achieve this,  we can borrow some ideas from Kit Fine's ``rigid embodiment'' and add primitive ordering relations to enrich the structure of mereological sums. }


In addition to  \textbf{N-Regions}, the quantum state amplitude structure will contain two primitive relations: 

\begin{itemize}
\item A two-place relation Amplitude--Geq ($\succeq_A$).
\item A three-place relation Amplitude--Sum ($S$). 
\end{itemize}

Interpretation:  $a \succeq_A b$ iff the amplitude of N-Region $a$ is greater than or equal to that of N-Region $b$; $S(a,b,c)$ iff the amplitude of N-Region $c$ is the sum of those of N-Regions $a$ and $b$.  

Define the following short-hand (all quantifiers below range over only N-Regions):

\begin{enumerate}
\item $a=_{A} b := a\succeq_{A} b  \text{ and  } b\succeq_{A} a. $
\item $a\succ_{A} b := a\succeq_{A} b  \text{ and not } b\succeq_{A} a. $

\end{enumerate}

Next, we can write down some axioms for Amplitude--Geq and Amplitude--Sum.\footnote{Compare with the axioms in Krantz et al. (1971) Defn.3.3:
Let $A$ be a nonempty set, $\succeq$ a binary relation on $A$, $B$ a nonempty subset of $A \times A$, and $\circ$ a binary function from $B$ into $A$. The quadruple $<A, \succeq, B, \circ>$ is an extensive structure with no essential maximum if the following axioms are satisfied for all $a,b,c\in A$:
\begin{enumerate}
\item $<A, \succeq>$ is a weak order. [This is translated as G1 and G2.]
\item If $(a,b)\in B$ and $(a\circ b, c) \in B$, then $(b,c)\in B, (a,  b\circ c)\in B$, and $(a\circ b)\circ c \succeq_{A} a\circ (b\circ c).$  [This is translated as S1.]
\item If $(a,c)\in B$ and $a \succeq b$, then $(c,b)\in B$, and $a\circ c \succeq  c\circ b.$ [This is translated as S2.]
\item If $a\succ b$,  then  $\exists d\in A$ s.t. $(b,d)\in B$ and $a \succeq b\circ d.$ [This is translated as S3.]
\item If $a\circ b = c$, then $c \succ a$. [This is translated as S4, but allowing N-Regions to have null amplitudes. The representation function will also be zero-valued at those regions.]
\item Every strictly bounded standard sequence is finite, where $a_1, ... , a_n, ...$ is a standard sequence if for $n=2,.., a_n = a_{n-1} \circ a_1$, and it is strictly bounded if for some $b\in A$ and for all $a_n$ in the sequence, $b\succ a_n$. [This is translated as S5. The translation uses the fact that Axiom 6 is equivalent to another formulation of the Archimedean axiom: $\{ n | na \text{ is defined and } b\succ na  \}$ is finite.]
\end{enumerate}
The complications in the nominalistic axioms come from the fact that there can be more than one N-Regions that are the Amplitude-Sum of two N-Regions: $\exists a,b,c,d$ s.t. $S(a,b,c) \wedge S(a,b,d) \wedge c\neq d$. However, in the proof for the representation and uniqueness theorems, we can easily overcome these complications by taking equivalence classes of equal amplitude and recover the amplitude addition function from the Amplitude-Sum relation. 
}  
Again, all quantifiers below range over only N-Regions.  $\forall a, b, c:$

\begin{enumerate}
\item[G1] (Connectedness) Either $a \succeq_{A} b$ or $b \succeq_{A} a$.
\item[G2] (Transitivity) If $a \succeq_{A} b$ and $b \succeq_{A} c$, then $a \succeq_{A} c$. 


\item[S1] (Associativity*) If $\exists x$ $S(a,b,x)$ and $\forall x'$ [if $S(a,b,x'))$ then $\exists y$  $S(x', c, y)$], then  $\exists z$  $S(b,c,z)$ and $\forall z'$ [if $S(b,c,z'))$ then $\exists w$  $S(a, z', w)$] 
and $\forall f, f', g, g' $ [if $S(a,b,f) \wedge S(f,c,f') \wedge S(b,c,g) \wedge S(a,g,g')$, then $f' \succeq_A g'$].



\item[S2] (Monotonicity*) If $ \exists x$ $S(a,c,x)$ and $a \succeq_{A} b$, then $ \exists y$  $S(c,b,y)$ and $\forall f, f'$ [if $S(a,c,f) \wedge S(c,b,f')$ then $f \succeq_{A}  f'$].


\item[S3] (Density) If $a\succ_{A} b$,  then $\exists d, x$  [$S(b,d, x)$ and $\forall f,$  if $S(b,x,f),$ then $a \succeq_{A}  f]$.

\item[S4] (Non-Negativity) If $S(a,b,c)$, then $c \succeq_{A} a$. 



\item[S5] (Archimedean Property) $\forall a_1, b$, if $\neg S(a_1, a_1, a_1)$ and $\neg S(b,b,b)$, then  $\exists a_1, a_2, ..., a_n$  s.t. $ b\succ_{A} a_n $ and $\forall a_i$  [if $b\succ_{A} a_i $, then $a_n \succeq_{A} a_i$], where $a_i$'s, if they exist, have the following properties: 
 $ S( a_1, a_1, a_2)$,  $ S( a_1, a_2, a_3)$, $ S( a_1, a_3, a_4)$, ...,  $ S( a_1, a_{n-1}, a_n)$.\footnote{S5 is  an infinitary sentence, as the quantifiers in the consequent should be understood as infinite disjunctions of quantified sentences. However, S5 can also be formulated with a stronger axiom called Dedekind Completeness, whose platonistic version says: 

\begin{description}
\item{\textbf{Dedekind Completeness.}} $\forall M, N \subset A$, if $\forall x\in M, \forall y \in N, y  \succ x$, then $\exists z \in A$ s.t. $\forall x\in M, z \succ x \text{ and } \forall y \in N, y  \succ z.$
\end{description}
The nominalistic translation can be done in two ways. We can introduce two levels of mereology so as to distinguish between regions of points and regions of regions of points. Alternatively, as Tom Donaldson, Jennifer Wang, and Gabriel Uzquiano suggest to me, perhaps one can make do with plural quantification in the following way. For example ( with $\propto$ for the logical predicate ``is one of'' ), here is one way to state the Dedekind Completeness with plural quantification: 

\begin{description}
\item{\textbf{Dedekind Completeness Nom Pl.}} $\forall mm, nn \in$ N-Regions, if $\forall x\propto mm, \forall y \propto nn, y  \succ x$, then there exists $z \in A$ s.t. $\forall x\propto mm, z \succ x \text{ and } \forall y \propto nn, y  \succ z.$
\end{description}
We only need the Archimedean property in the proof. Since Dedekind Completeness is stronger, the proof in Krantz et al. (1971), pp. 84-87  can still go through if we assume Dedekind Completeness Nom Pl.
Such strenghthening of S5 has the virtue of avoiding the infinitary sentences in S5. Note: this is the point where we have to trade off certain nice features of first-order logic and standard mereology with the desiderata of the intrinsic and nominalistic account. (I have no problem with infinitary sentences in S5. But one is free to choose instead to use plural quantification to formulate the last axiom as Dedekind Completeness Nom Pl.) This is related to Field's worry in \emph{Science Without Numbers}, Ch. 9, ``Logic and Ontology.''}


\end{enumerate}
Since these axioms are the nominalistic translations of a platonistic structure in Krantz et al. (Defn. 3.3), we can formulate the representation and uniqueness theorems for the amplitude structure as follows:

\begin{theorem}[Amplitude Representation Theorem]
  <N-Regions, Amplitude--Geq, \text{   } Amplitude--Sum> satisfies axioms (G1)---(G2) and (S1)---(S5), only if there is a function $R: \text{ N-Regions} \rightarrow   \{0\} \cup \mathbb{R}^{+}$ such that  $\forall a, b \in \text{N-Regions}$:
\begin{enumerate}
\item $a\succeq_{A} b \Leftrightarrow R(a) \geq R(b)$;
\item If $\exists x$ s.t. $S(a,b,x)$, then $\forall c$ [if $S(a,b,c)$ then $R(c)=R(a) +R(b)$].
\end{enumerate}

\end{theorem}

\begin{theorem}[Amplitude Uniqueness Theorem]

If another function $R'$ satisfies the conditions on the RHS of the Amplitude Representation Theorem, then there exists a real number $\alpha > 0$ such that for all nonmaximal element $a\in $ N-Regions, $R'(a) = \alpha R(a). $

\end{theorem}

Proofs: See  \cite{KrantzLuceSuppesTversky}, Sections 3.4.3, 3.5, pp. 84-87. Note: Krantz et al. use an addition function $\circ$, while we use a sum relation $S(x,y,z)$, because we allow there to be distinct N-Regions that have the same amplitude. Nevertheless, we can use a standard technique to adapt their proof: we can simply take the equivalence classes N-Regions / $=_A$, where $a =_A b$ if $a\succeq_{A} b \wedge b\succeq_{A} a$, on which we can define an addition function with the Amplitude-Sum relation.

The representation theorem suggests that the intrinsic structure of Amplitude-Geq and Amplitude-Sum guarantees the existence of a faithful representation function. But the intrinsic structure makes no essential quantification over numbers, functions, sets, or matrices. The uniqueness theorem explains why the gauge degrees of freedom are the positive multiplication transformations and no further, i.e. why the amplitude function is unique up to a positive normalization constant. 

\subsection{Quantum State Phase}

The phase part of the quantum state is (like angles on a plane) of the periodic scale, i.e. the intrinsic physical structure should be invariant under overall phase transformations $$\theta \rightarrow \theta + \phi \text{ mod } 2\pi.$$ 
We would like something of the form of a ``difference structure.'' But we know that according to standard formalism, just the absolute values of the differences would not be enough, for time reversal on the quantum state is implemented by taking the complex conjugation of the wave function, which is an operation that leaves the absolute values of the differences unchanged. So we will try to construct a signed difference structure such that standard operations on the wave function are faithfully preserved.\footnote{Thanks to Sheldon Goldstein for helpful discussions about this point. David Wallace points out
(p.c.) that it might be a virtue of the nominalistic theory to display the following choice-point: one can imagine an axiomatization of quantum state phase that involves only absolute phase differences. This would require thinking more deeply about the relationship between quantum phases and temporal structure, as well as a new mathematical axiomatization of the absolute difference structure for phase.}

We will once again start with \textbf{N-Regions}, the collection of all  regions that contain exactly $N$ simultaneous space-time points. 

The intrinsic structure of phase consists in two primitive relations: 
\begin{itemize}
\item A three-place relation Phase--Clockwise--Betweenness ($C_{P}$), 
\item A four-place relation Phase--Congruence ($\sim_{P}$). 
\end{itemize}

Interpretation:  $C_{P} (a,b,c)$ iff the phase of N-Region $b$ is clock-wise between those of N-Regions $a$ and $c$ (this relation realizes the intuitive idea that 3 o'clock is clock-wise between 1 o'clock and 6 o'clock, but 3 o'clock is not clock-wise between 6 o'clock and 1 o'clock); $ ab \sim_{P} cd$ iff the signed phase difference between N-Regions $a$ and $b$ is the same as that between N-Regions $c$ and $d$. 




The intrinsic structures of Phase--Clockwise--Betweenness and Phase--Congruence satisfy the following axioms for what I shall call a ``periodic difference structure'':

All quantifiers below range over only N-Regions.
$\forall a, b, c, d, e, f$:

\begin{enumerate}
\item[C1] At least one of $C_{P}(a,b,c)$ and $C_{P}(a,c,b)$ holds; if $a,b,c$ are pair-wise distinct, then exactly one of $C_{P}(a,b,c)$ and $C_{P}(a,c,b)$ holds.
\item[C2] If $C_{P}(a,b,c)$ and $C_{P}(a,c,d)$, then $C_{P}(a,b,d)$; if $C_{P}(a,b,c)$, then $C_{P}(b,c,a)$. 
\item[K1] $ab \sim_P ab$. 
\item[K2] $ab \sim_P cd \Leftrightarrow cd \sim_P ab \Leftrightarrow ba \sim_P dc \Leftrightarrow ac \sim_P bd$. 
\item[K3] If $ab \sim_P cd$ and $cd \sim_P ef$, then $ab \sim_P ef$. 



\item[K4] $\exists h, cb\sim_P ah$; if $C_P (a,b, c)$, then $\exists d',d''$ s.t.  $ba\sim_P d'c,$  $ca\sim_P d''b$; $\exists p, q, C_P (a,q,b), C_P (a,b, p),$ $ap \sim_P pb, bq \sim_P qa.$ 

\item[K5] $ab\sim_P cd \Leftrightarrow  [ \forall e, fd\sim_P ae \Leftrightarrow fc\sim_P be].$  

\item[K6] $\forall e,f,g,h,$ if $fc\sim_P be$ and $gb\sim_P ae$, then $[hf\sim_P ae \Leftrightarrow hc\sim_P ge]. $  

\item[K7] If $ C_P (a,b,c)$, then $\forall e,d,a',b', c'$ [if $a'd \sim_P ae, b'd \sim_Pbe, c'd \sim_P ce,$ then $C(a',b',c')$].



\item[K8] (Archimedean Property) $\forall a, a_1, b_1$, if $C_P (a,a_1, b_1)$, then $\exists a_1, a_2, ..., a_n, b_1, b_2, ..., b_n, c_1, ...c_m$ such that $C_P (a,a_1, a_n)$ and $C_P (a,b_n, b_1)$, where $a_n a_{n-1} \sim_P a_{n-1} a_{n-2} \sim_P ... \sim_P a_{1} a_{2}$ and $b_n b_{n-1} \sim_P b_{n-1} b_{n-2} \sim_P ... \sim_P b_{1} b_{2}$, and that $a_1 b_1 \sim_P b_1 c_1 \sim_P ... \sim_P c_n a_1$.\footnote{Here it might again be desirable to avoid the infinitary sentences / axiom schema by using plural quantification. See the footnote on Axiom S5.} 

\end{enumerate}

Axiom (K4) contains several existence assumptions. But such assumptions are justified for a nominalistic quantum theory. We can see this from the structure of the platonistic quantum theory. Thanks to the Schr\"odinger dynamics, the wave function will spread out continuously over space and time, which will ensure the richness in the phase structure. 

With some work, we can prove the following representation and uniqueness theorems: 

\begin{theorem}[Phase Representation Theorem]
 If  $<$ N-Regions, Phase--Clockwise--Betweenness, Phase--Congruence$>$ is a periodic difference structure, i.e. satisfies axioms (C1)---(C2) and (K1)---(K8), then for any real number $k > 0$,  there is a function $f: \text{ N-Regions } \rightarrow [0, k)$ such that  $\forall a, b, c, d \in $ N-Regions:
\begin{enumerate}
\item $C_{P} (c,b,a) \Leftrightarrow f(a) \geq f(b) \geq f(c) \text{ or } f(c) \geq f(a) \geq f(b)  \text{ or } f(b)\geq f(c)\geq f(a) $;
\item $ab \sim_{P} cd \Leftrightarrow f(a)-f(b) = f(c)-f(d)$ (mod $k$).
\end{enumerate}
\end{theorem}

\begin{theorem}[Phase Uniqueness Theorem]

If another function $f'$ satisfies the conditions on the RHS of the Phase Representation Theorem, then there exists a real number $\beta $ such that for all element $a\in $ N-Regions , $f'(a) =  f(a) +\beta $ (mod $k$).

\end{theorem}
Proofs: see Appendix. 

Again, the representation theorem suggests that the intrinsic structure of Phase--Clockwise--Betweenness and Phase--Congruence guarantees the existence of a faithful representation function of phase. But the intrinsic structure makes no essential quantification over numbers, functions, sets, or matrices. The uniqueness theorem explains why the gauge degrees of freedom are the overall phase transformations and no further, i.e. why the phase function is unique up to an additive constant. 

Therefore, we have written down an intrinsic and nominalistic theory of the quantum state, consisting in merely four relations on the regions of physical space-time: Amplitude-Sum, Amplitude-Geq, Phase-Congruence, and Phase-Clockwise-Betweenness. As mentioned earlier but evident now, the present account of the quantum state has several desirable features:  (1) it does not refer to any abstract mathematical objects such as complex numbers, (2) it is free from the usual arbitrary conventions in the wave function representation, and (3) it explains why the quantum state has its amplitude and phase degrees of freedom.

\subsection{Comparisons with Balaguer's Account}

Let us briefly compare my account with Mark Balaguer's account (1996) of the nominalization of quantum mechanics.

Balaguer's account follows Malament's suggestion of nominalizing quantum mechanics by taking seriously the Hilbert space structure and the representation of ``quantum events'' with closed subspaces of Hilbert spaces. Following orthodox textbook presentation of quantum mechanics, he suggests that we take as primitives the \emph{propensities} of quantum systems as analogous to probabilities of quantum experimental outcomes. 
\begin{quote}
I begin by recalling that each quantum state can be thought of as a function from events $(A, \Delta)$ to probabilities, i.e., to $[0,1]$. Thus, each quantum state specifies a set of ordered pairs $<(A, \Delta), r>$. The next thing to notice is that each such ordered pair determines a propensity property of quantum systems, namely, an $r-$strengthed propensity to yield a value in $\Delta$ for a measurement of $A$. We can denote this propensity with ``$(A, \Delta, r)$''.  (Balaguer, 1996, p.218.)
\end{quote}
Balaguer suggests that the propensities are ``nominalistically kosher.'' By interpreting the Hilbert space structures as propensities instead of propositions, Balaguer makes some progress in the direction of making quantum mechanics ``more nominalistic.''

However, Balaguer's account faces a problem---it is not clear how Balaguer's account relates to any mainstream realist interpretation of quantum mechanics. This is because the realist interpretations---Bohmian Mechanics, GRW spontaneous collapse theories, and Everettian Quantum Mechanics---crucially involve the quantum state represented by a wave function, not a function from events to probabilities.\footnote{See \cite{bueno2003possible} for a discussion about the conflicts between Balaguer's account and the modal interpretation of QM.} And once we add the wave function (perhaps in the nominalistic form introduced in this paper), the probabilities can be calculated (via the Born rule) from the wave function itself, which makes primitive propensities redundant. If Balaguer's account is based on orthodox quantum mechanics, then it would suffer from the dependence on vague notions such as ``measurement,'' ``observation,'' and ``observables,'' which should have no place in the fundamental ontology or dynamics of a physical theory.\footnote{Bell (1989), ``Against `Measurement,' '' pp. 215-16.}  


\section{``Wave Function Realism''}

The intrinsic and nominalistic account of the quantum state provides a natural response to some of the standard objections to ``wave function realism.''\footnote{``Wave function realists,'' such as David Albert, Barry Loewer, Alyssa Ney, and Jill North, maintain that the fundamental physical space for a quantum world is 3N-dimensional. In contrast, primitive ontologists, such as Valia Allori, Detlef D\"urr, Sheldon Goldstein, Tim Maudlin, Roderich Tumulka, and Nino Zanghi, argue that the fundamental physical space is 3-dimensional.} 
According to David Albert (1996), realism about the wave function naturally commits one to accept that the wave function is a physical field defined on a fundamentally 3N-dimensional wave function space. Tim \cite{maudlin2013nature} criticizes Albert's view partly on the ground that such  ``naive'' realism would commit one to take as fundamental the  gauge degrees of freedom such as the absolute values of the amplitude and the phase, leaving empirically equivalent formulations as metaphysically distinct. This ``naive'' realism is inconsistent with the physicists' attitude of understanding the Hilbert space projectively and thinking of the quantum state as an equivalence class of wave functions ($\psi \sim R e^{i\theta} \psi$).  If a defender of wave function realism were to take the physicists' attitude, says the opponent, it would be much less natural to think of the wave function as really a physical \emph{field}, as something that assigns physical properties to each point in the 3N-dimensional space. Defenders of wave function realism have largely responded by biting the bullet and accepting the costs.

But the situation changes given the present account of the quantum state. Given the intrinsic theory of the quantum state, one can be realist about the quantum state by being realist about the four intrinsic relations underlying the mathematical and gauge-dependent description of the wave function. The intrinsic relations are invariant under the gauge transformations. Regardless of whether one believes in a fundamentally high-dimensional space or a fundamentally low-dimensional space, the intrinsic account will recover the wave function unique up to the gauge transformations ($\psi \sim R e^{i\theta} \psi$). 

I should emphasize that my intrinsic account of the wave function is essentially a version of comparativism about quantities. As such, it should be distinguished from eliminitivism about quantities.  Just as a comparativist about mass does not eliminate mass facts but ground them in comparative mass relations, my approach does not eliminate wave function facts but ground them in comparative amplitude and phase relations. My account does not in the least suggest any anti-realism about the wave function.\footnote{ Thanks to David Glick for suggesting that I make this clear.}

Therefore, my account removes a major obstacle for wave function realism.  One can use the intrinsic account of the quantum state to identify two field-like entities on the configuration space (by thinking of the N-Regions as points in the 3N-dimensional space) without committing to the excess structure of absolute amplitude and overall phase.\footnote{ Unsurprisingly, the present account  also provides some new arsenal for the defenders of the fundamental 3-dimensional space. The intrinsic account of the quantum state fleshes out some details in the multi-field proposal.}

\section{Conclusion}

There are many \emph{prima facie} reasons for doubting that we can ever find an intrinsic and nominalistic theory of quantum mechanics. However, in this paper, we have offered an intrinsic and nominalistic account of the quantum state, consisting in  four relations on  regions of physical space: 
\begin{enumerate}
\item Amplitude-Sum ($S$), 
\item Amplitude-Geq ($\succeq_A$), 
\item Phase-Congruence ($\sim_P$), 
\item Phase-Clockwise-Betweenness ($C_P$). 
\end{enumerate}
This account, I believe, offers a deeper understanding of the nature of the quantum state that at the very least complements that of the standard account. By removing the references to mathematical objects, our account of the quantum state provides a framework for nominalizing quantum mechanics. By excising superfluous structure such as overall phase, it reveals the intrinsic structure postulated by quantum mechanics. Here we have focused on the universal quantum state. As the origin of quantum non-locality and randomness, the universal wave function has no classical counterpart and seems to resist an intrinsic and nominalistic treatment. With the focus on the universal quantum state,  our account still leaves room for including additional ontologies of particles, fields, mass densities supplied by specific solutions to the quantum measurement problem such as BM, GRWm, and GRWf; these additional ontologies are readily nominalizable. 

Let us anticipate some directions for future research. First, the intrinsic structure of the quantum state at different times is constrained by the quantum dynamics. In the platonistic theory, the dynamics is described by the Schr\"odinger equation. To nominalize the dynamics, we can decompose  the Schr\"odinger equation into two equations, in terms of amplitude and gradient of phase of the wave function. The key would be to codify the differential operations (which Field has done for Newtonian Gravitation Theory) in such a way to be compatible with our phase and amplitude relations. Second, we have described how to think of the quantum state for a system with constant number of particles. How should we extend this account to accommodate particle creation and annihilation in quantum field theories? I think the best way to answer that question would be to think carefully about the ontology of a quantum field theory. A possible interpretation is to think of the quantum state of a variable number of particles as being represented by a complex valued function whose domain is $\bigcup_{N=0}^{\infty} \mathbb{R}^{3N}$---the union of all configuration spaces (of different number of particles). In that case, the extension of our theory would be easy:  (1) keep the axioms as they are and (2) let the quantifiers range over $K$-regions, where the integer $K$ ranges from zero to infinity. Third, we have  considered quantum states for spinless systems. A possible way to extend the present account to accommodate spinorial degrees of freedom would be to use two comparative relations for each complex number assigned by the wave function. That strategy is conceptually similar to the situation in the present account. But it is certainly not the only strategy, especially considering the gauge degrees of freedom in the spin space. Fourth, as we have learned from debates about the relational theories of motion and the comparative theories of quantities, there is always the possibility of a theory becoming indeterministic when drawing from only comparative predicates without fixing an absolute scale.\footnote{See  \cite{dasgupta2013absolutism, baker2014some, martens2016against}, and Field (2016), preface to the second edition, pp. 41-44. } It would be interesting to investigate whether similar problems of indeterminism arise in our comparative theory of the quantum state.  Finally, the formal results obtained for the periodic difference structure could be useful for further investigation into the metaphysics of phase.

The nature of the quantum state is the origin of many deeply puzzling features of a quantum world. It is especially challenging given realist commitments. I hope that the account discussed in this paper makes some progress towards a better understanding of it.

\section*{Acknowledgement}
Many thanks to Cian Dorr, Hartry Field, and David Malament for their encouragements, criticisms, and extensive feedback on this project. For helpful discussions, I'd like to thank David Albert,  Andrew Bacon, Mark Balaguer, Jeff Barrett, Thomas Barrett, Rhys Borchert, Adam Caulton, Shamik Dasgupta, Tom Donaldson, Joshua Gert, Sheldon Goldstein, Veronica Gomez, Jeremy Goodman, Richard Healey, James Ladyman, Barry Loewer, Tim Maudlin, Richard Pettigrew, Carlo Rovelli, Jeff Russell, Charles Sebens, Ward Struyve, David Wallace, Jennifer Wang, Nino Zangh\`i, audiences at the UC Irvine philosophy of physics reading group, Rutgers metaphysics reading group, Bristol Center for Science and Philosophy, the 2016 Foundations of Physics Conference at LSE, the 3rd conference of the Society of Metaphysics of Science at Fordham,  the 4th Black Forest Summer School in Philosophy of Physics in Saig, Germany, and a symposium at the 2018 Pacific APA meeting in San Diego. I'd especially like to thank Liam Kofi Bright, Kenny Easwaran, David Glick, Jill North, Andrea Oldofredi, Bryan Roberts, Ted Sider, and Isaac Wilhelm  for their insightful and detailed written comments.


\section*{Appendix: Proofs of Theorems 3.3 and 3.4.}

\textbf{Step 1.} We begin by enriching $<$ N-Regions, Phase--Clockwise--Betweenness, Phase--Congruence$>$ with some additional structures. 

First, to simplify the notations, let us think of N-Regions as a set of regions, and let us now only consider $\Omega :=$ N-Regions /  $=_P$, the set of ``equal phase'' equivalence classes by quotienting out $=_P$. ($a=_P b $ if they form phase intervals the same way: $\forall c \in S, ac \sim_P bc$.) 

Second, we fix an arbitrary $A_0 \in \Omega $ to be the ``zero phase equivalence class.'' 

Third, we define a non-inclusive relation $C$ on $\Omega$ according to $C_P$ on N-Regions. ($\forall A, B, C \in \Omega $, $C(A,B,C)$ iff $A, B, C$ are pairwise distinct and $\forall a\in A, \forall b \in B, \forall c \in C$, $C(a,b,c)$.) 


Fourth, we define an addition function $\circ:$ $\Omega\times \Omega \rightarrow \Omega$.  $\forall A, B \in \Omega$, $C=A\circ B$ is the unique element in $\Omega$ such that $CB \sim AA_0$, which is guaranteed to exist by (K4) and provably unique as elements in $\Omega$ form a partition over N-Regions.

\textbf{Step 2.} We show that the enriched structure $< \Omega, \circ, C>$ with identity element $A_0$ satisfies the axioms for a periodic extensive structure defined in \cite{luce1971periodic}. 

\textbf{Axiom 0}. $<\Omega, \circ>$ is an Abelian semigroup. 

First, we show that $\circ$ is closed: $\forall A, B \in \Omega$, $A\circ B \in \Omega$. 

This follows from (K4). 

Second, we show that $\circ$ is associative: $\forall A, B, C \in \Omega$, $A\circ(B\circ C) = (A\circ B) \circ C$. 

This follows from (K6).

Third, we show that $\circ$ is commutative: $\forall A, B \in \Omega$, $A\circ B = B\circ A$. 

This follows from (K2).

$\forall A, B, C, D\in \Omega$:

\textbf{Axiom 1}. Exactly one of $C(A,B,C)$ or $C(A,C,B)$ holds.

This follows from C1. 

\textbf{Axiom 2}. $C(A,B,C)$ implies $C(B,C,A)$.

This follows from C2.

\textbf{Axiom 3}. $C(A,B,C)$ and $C(A,C,D)$ implies $C(A,B,D)$.

This follows from C2.

\textbf{Axiom 4}. $C(A,B,C)$ implies $C(A\circ D,B\circ D,C\circ D)$ and $C(D\circ A, D\circ B, D\circ C)$.

This follows from (K7).





\textbf{Axiom 5}. If $C(A_0,A,B)$, then there exists a positive integer $n$ such that $C(A_0,A,nA)$ and $C(A_0,nB,B)$. 

This follows from (K8). 

Therefore, the enriched structure $< \Omega, \circ, C>$ with identity element $A_0$ satisfies the axioms for a periodic extensive structure defined in \cite{luce1971periodic}.

\textbf{Step 3.} We use the homomorphisms in \cite{luce1971periodic} to find the homomorphisms for < N-Regions, Phase--Clockwise--Betweenness, Phase--Congruence>.  

Since $< \Omega, \circ, C>$  satisfy the axioms for a periodic structure, Corollary in \cite{luce1971periodic} says that for any real $K>0$, there is a unique function $\phi$ from $\Omega$ into $[0,K)$ s.t. $\forall A,B,C \in \Omega$:

\begin{enumerate}
\item $C (C,B,A)$ $\Leftrightarrow$ $\phi(A) > \phi(B) > \phi(C)$  or $\phi(C) > \phi(A) > \phi(B) $ or $ \phi(B)> \phi(C)> \phi(A) $;
\item  $\phi(A\circ B) = \phi(A) + \phi(B)$ (mod $K$);
\item $\phi(A_0) = 0$.
\end{enumerate}

Now, we define $f:$ N-Regions $\rightarrow [0,K)$ as follows: $f(a) = \phi (A)$, where $a\in A$. So we have 
 $C_{P} (c,b,a) \Leftrightarrow f(a) \geq f(b) \geq f(c) \text{ or } f(c) \geq f(a) \geq f(b)  \text{ or } f(b)\geq f(c)\geq f(a) $.

We can also define $\psi:$ N-Regions $\times$ N-Regions $\rightarrow [0,K)$ as follows: $\psi(a,b) = \phi(A) - \phi(B)$ (mod $K$), where $a\in A$ and $b\in B$. Hence, $\forall a,b\in $ N-Regions, $\psi(a,b) = f(a) - f(b)$  (mod $K$).

Moreover, given (K5), $\forall a\in A, b\in B, c\in C, d\in D,$ $ab \sim_{P} cd$ 

$\Leftrightarrow$ $AB \sim CD$

$\Leftrightarrow$ $A \circ D = B \circ C$ 

$\Leftrightarrow$ $\phi(A \circ D) = \phi(B \circ C)$ 

$\Leftrightarrow$ $\phi(A) + \phi(D) = \phi(B) + \phi(C)$ (mod $K$) 

$\Leftrightarrow$ $\forall a\in A, b\in B, c\in C, d\in D,$ $f(a) + f(d) = f(b) + f(c)$ (mod $K$)

$\Leftrightarrow$ $\forall a\in A, b\in B, c\in C, d\in D,$ $f(a) - f(b) = f(c) - f(d)$ (mod $K$)

Therefore, we have demonstrated the existence of homomorphisms. 

\textbf{Step 4.} We prove the uniqueness theorem. 

If another function $f': \text{ N-Regions } \rightarrow [0,K)$ with the same properties exists, then it satisfies the same homomorphism conditions. Suppose (for \emph{reductio}) $f'$ and $f$ differ by more than a constant mod K. 

Let $D(a,b)$ by the function that measures the differences between $f$ and $f'$ on N-regions:
 $$ D(a,b) = [f(a) - f(b) ] - [f'(a) - f'(b)] \text{ mod } K.$$ 

Without loss of generality, let us suppose that there exist two regions $x,y$ where $D(x,y) \neq 0$. By (K8), there will be a sequence of pairs of regions that are phase-congruent to $x,y$: $xy \sim_P ya_1 \sim_P a_1 a_2 \sim_P ... \sim_P a_n x$. Since by assumption both $f$ and $f'$ preserve the structure of phase-congruence, we have (mod K): 
$$f(x)-f(y) = f(y)-f(a_1 ) = ... = f(a_n ) - f(x),$$
$$f'(x)-f'(y) = f'(y)-f'(a_1 ) = ... = f'(a_n ) - f'(x).$$
Consequently:
$$D(x,y) = [f(x) - f(y)] -  [f'(x) - f'(y)] = D(y, a_1 ) = ... = D(a_n , x) $$
Hence, since the above $D$'s are not zero,  they must be either all positive or all negative. If they are all positive, then the sum of them will be positive:
$$D(x,y) +  D(y, a_1 ) + ... + D(a_n , x) > 0$$
 However, expanding them in terms of $f$ and $f'$ we have a telescoping sum:
$$[f(x) - f(y)] -  [f'(x) - f'(y)] + [f(y) - f(a_1 )] -  [f'(y) - f'(a_1 )] + ... + [f(a_n ) - f(x)] -  [f'(a_n ) - f'(x)] =0.$$
Contradiction. The same argument works for the case when all the $D$'s are negative. 

Therefore, $D(x,y) = 0$ for all N-regions. Let $a_0$ be where $f$ assigns zero. Then 
$$f'(a)-f'(a_0) \text{ mod } K  = f(a)-f(a_0) \text{ mod } K = f(a),$$ 
which entails that 
$$f'(a) = f(a) + \beta \text{ mod } K,$$ 
with the constant $\beta = f'(a_0)$.  QED. \\



\bibliography{test}

\begin{thebibliography}{}

\bibitem[Albert, 1996]{AlbertEQM}
Albert, D.~Z. (1996).
\newblock Elementary quantum metaphysics.
\newblock In Cushing, J.~T., Fine, A., and Goldstein, S., editors, {\em Bohmian
  Mechanics and Quantum Theory: An Appraisal}, pages 277--84. Kluwer Academic
  Publishers, Dordrecht.

\bibitem[Allori, 2013]{allori2013primitive}
Allori, V. (2013).
\newblock Primitive ontology and the structure of fundamental physical
  theories.
\newblock {\em The Wave Function: Essays on the Metaphysics of Quantum
  Mechanics}, pages 58--75.

\bibitem[Arntzenius and Dorr, 2011]{ArntzeniusDorrCG}
Arntzenius, F. and Dorr, C. (2011).
\newblock Calculus as geometry.
\newblock Chapter 8 of Frank Arntzenius, \emph{Space, Time, and Stuff}. Oxford:
  Oxford University Press.

\bibitem[Baker, 2014]{baker2014some}
Baker, D.~J. (2014).
\newblock Some consequences of physics for the comparative metaphysics of
  quantity.
\newblock {\em PhilSci-Archive}.

\bibitem[Bell, 2004]{bell2004speakable}
Bell, J.~S. (2004).
\newblock {\em Speakable and Unspeakable in Quantum Physics: Collected Papers
  on Quantum Philosophy}.
\newblock Cambridge University Press.

\bibitem[Belot, 2012]{belot2012quantum}
Belot, G. (2012).
\newblock Quantum states for primitive ontologists.
\newblock {\em European Journal for Philosophy of Science}, 2(1):67--83.

\bibitem[Bhogal and Perry, 2015]{bhogal2015humean}
Bhogal, H. and Perry, Z. (2015).
\newblock What the humean should say about entanglement.
\newblock {\em No{\^u}s}.

\bibitem[Bueno, 2003]{bueno2003possible}
Bueno, O. (2003).
\newblock Is it possible to nominalize quantum mechanics?
\newblock {\em Philosophy of Science}, 70(5):1424--1436.

\bibitem[Bueno, 2014]{sep-nominalism-mathematics}
Bueno, O. (2014).
\newblock Nominalism in the philosophy of mathematics.
\newblock In Zalta, E.~N., editor, {\em The Stanford Encyclopedia of
  Philosophy}. Metaphysics Research Lab, Stanford University, spring 2014
  edition.

\bibitem[Chen, 2017]{ChenOurFund}
Chen, E.~K. (2017).
\newblock Our fundamental physical space: An essay on the metaphysics of the
  wave function.
\newblock {\em Journal of Philosophy}, 114: 7.

\bibitem[Colyvan, 2010]{colyvan2010there}
Colyvan, M. (2010).
\newblock There is no easy road to nominalism.
\newblock {\em Mind}, 119(474):285--306.

\bibitem[Colyvan, 2015]{sep-mathphil-indis}
Colyvan, M. (2015).
\newblock Indispensability arguments in the philosophy of mathematics.
\newblock In Zalta, E.~N., editor, {\em The Stanford Encyclopedia of
  Philosophy}. Metaphysics Research Lab, Stanford University, spring 2015
  edition.

\bibitem[Dasgupta, 2013]{dasgupta2013absolutism}
Dasgupta, S. (2013).
\newblock Absolutism vs comparativism about quantity.
\newblock {\em Oxford Studies in Metaphysics}, 8:105--148.

\bibitem[D{\"u}rr et~al., 2012]{durr2012quantum}
D{\"u}rr, D., Goldstein, S., and Zangh\`i, N. (2012).
\newblock {\em Quantum physics without quantum philosophy}.
\newblock Springer Science \& Business Media.

\bibitem[Field, 1980]{FieldSWN}
Field, H. (1980).
\newblock {\em Science Without Numbers}.
\newblock Blackwell, Oxford.

\bibitem[Field, 2016]{field2016science}
Field, H. (2016).
\newblock {\em Science without numbers}.
\newblock Oxford University Press, 2nd edition.

\bibitem[Forrest, 1988]{forrest1988quantum}
Forrest, P. (1988).
\newblock {\em Quantum metaphysics}.
\newblock Blackwell Publisher.

\bibitem[Goldstein and Zangh\`i, 2013]{goldstein2013reality}
Goldstein, S. and Zangh\`i, N. (2013).
\newblock Reality and the role of the wave function in quantum theory.
\newblock {\em The wave function: Essays on the metaphysics of quantum
  mechanics}, pages 91--109.

\bibitem[Hubert and Romano, 2018]{Hubert2018}
Hubert, M. and Romano, D. (2018).
\newblock The wave-function as a multi-field.
\newblock {\em European Journal for Philosophy of Science}, 8(3):521--537.

\bibitem[Krantz et~al., 1971]{KrantzLuceSuppesTversky}
Krantz, D.~H., Luce, R.~D., Suppes, P., and Tversky, A. (1971).
\newblock {\em Foundations of Measurement: Volume I, Additive and Polynomial
  Representations}.
\newblock Academic Press, New York.

\bibitem[Loewer, 1996]{LoewerHS}
Loewer, B. (1996).
\newblock Humean supervenience.
\newblock {\em Philosophical Topics}, 24:101--27.

\bibitem[Luce, 1971]{luce1971periodic}
Luce, R.~D. (1971).
\newblock Periodic extensive measurement.
\newblock {\em Compositio Mathematica}, 23(2):189--198.

\bibitem[Malament, 1982]{MalamentFR}
Malament, D. (1982).
\newblock Review of {H}artry {F}ield, \emph{Science without Numbers}.
\newblock {\em Journal of Philosophy}, 79:523--34.

\bibitem[Martens, 2016]{martens2016against}
Martens, N. (2016).
\newblock Against reducing newtonian mass to kinematical quantities.
\newblock {\em PhilSci-Archive}.

\bibitem[Maudlin, 2013]{maudlin2013nature}
Maudlin, T. (2013).
\newblock The nature of the quantum state.
\newblock {\em The wave function: Essays on the metaphysics of quantum
  mechanics}, pages 126--53.

\bibitem[Miller, 2013]{miller2013quantum}
Miller, E. (2013).
\newblock Quantum entanglement, bohmian mechanics, and humean supervenience.
\newblock {\em Australasian Journal of Philosophy}, (ahead-of-print):1--17.

\bibitem[Ney, 2012]{NeySOTDQU}
Ney, A. (2012).
\newblock The status of our ordinary three dimensions in a quantum universe.
\newblock {\em No\^{u}s}, 46:525--60.

\bibitem[Ney and Albert, 2013]{ney2013wave}
Ney, A. and Albert, D.~Z. (2013).
\newblock {\em The wave function: Essays on the metaphysics of quantum
  mechanics}.
\newblock Oxford University Press.

\bibitem[Norsen, 2017]{norsen2017}
Norsen, T. (2017).
\newblock {\em Foundations of Quantum Mechanics: An Exploration of the Physical
  Meaning of Quantum Theory}.
\newblock Springer.

\bibitem[North, 2009]{north2009structure}
North, J. (2009).
\newblock The ``structure" of physics: A case study.
\newblock {\em The Journal of Philosophy}, 106(2):57--88.

\bibitem[North, 2013]{NorthSQW}
North, J. (2013).
\newblock The structure of a quantum world.
\newblock In Albert, D.~Z. and Ney, A., editors, {\em The Wavefunction: Essays
  in the Metaphysics of Quantum Mechanics}. Oxford University Press, Oxford.
\newblock Forthcoming.

\bibitem[Putnam, 1971]{PutnamPL}
Putnam, H. (1971).
\newblock {\em Philosophy of Logic}.
\newblock Harper {\&} Row, New York.

\bibitem[Varzi, 2016]{sep-mereology}
Varzi, A. (2016).
\newblock Mereology.
\newblock In Zalta, E.~N., editor, {\em The Stanford Encyclopedia of
  Philosophy}. Metaphysics Research Lab, Stanford University, winter 2016
  edition.

\bibitem[Wallace and Timpson, 2010]{wallace2010quantum}
Wallace, D. and Timpson, C.~G. (2010).
\newblock Quantum mechanics on spacetime {I}: Spacetime state realism.
\newblock {\em The British Journal for the Philosophy of Science},
  61(4):697--727.

\end{thebibliography}


\end{document}